\documentclass{ws-procs975x65}
\newcommand {\beq}{\begin{equation}}
\newcommand {\eeq}{\end{equation}}
\newcommand {\bea}{\begin{eqnarray}}
\newcommand {\eea}{\end{eqnarray}}
\newcommand {\nn}{\nonumber \\}
\newcommand {\e}{{\rm e}}       %090315  erase No.2

\newcommand {\m}{\mu}
\newcommand {\n}{\nu}
\newcommand {\pl}{\partial}

\newcommand {\al}{\alpha}
\newcommand {\be}{\beta}

\newcommand {\la}{\lambda}
\newcommand {\La}{\Lambda}

\newcommand {\om}{\omega}

\newcommand {\na}{\nabla}
\newcommand {\del}  {\delta}

\newcommand {\mn}{{\mu\nu}}

\newcommand {\half}{ {\frac{1}{2}} }     %090315  erse No.3 

\newcommand {\fourth} {\frac{1}{4} }
\newcommand {\Ecal}{{\cal E}}
\newcommand {\Fcal}{{\cal F}}
\newcommand {\Lcal}{{\cal L}}

\newcommand {\Pcal}{{\cal P}}
\newcommand {\Dcal}{{\cal D}}

\newcommand {\omvec}{{\vec \omega}}
\newcommand {\ptil} {{\tilde p}}
\newcommand {\ktil} {{\tilde k}}

\newcommand {\omtil}{{\tilde \omega}}
\newcommand {\Lhat}{{\hat L}}

\newcommand {\bfZ} {{\bf Z}}
\newcommand {\K}{{\bf K}}
\newcommand {\I}{{\bf I}}

\newcommand {\intfx} {{\int d^4x}}

\newcommand {\intxz} {{\int d^4xdz}}

\newcommand {\intp} {{\int \frac{d^4p}{(2\pi)^4}}}

\newcommand {\intt} {{\int_{0}^{\infty}\frac{dt}{t}}}
\newcommand {\change} {\leftrightarrow}
\newcommand {\ra} {\rightarrow}

\newcommand {\pr}   {{\quad .}}
\newcommand {\com}  {{\quad ,}}
\newcommand {\q}    {\quad}

\newcommand {\NP}   {Nucl.Phys.}

\newcommand {\PR}   {Phys.Rev.}
\newcommand {\PRL}   {Phys.Rev.Lett.}

\newcommand {\PTP}  {Prog.Theor.Phys.}

\newcommand {\CQG}  {Class.Quantum.Grav.}

\newcommand {\Pla} {\frac{{\tilde p}}{\omega}}

\newcommand {\Tev} {\frac{{\tilde p}}{T}}

\begin{document}

\title{Casimir Energy of AdS5 Electromagnetism and 
Cosmological Constant Problem}

\author{S. Ichinose}

\address{
Laboratory of Physics, School of Food and Nutritional Sciences, 
University of Shizuoka,\\
Yada 52-1, Shizuoka 422-8526, Japan\\
$^*$E-mail: ichinose@u-shizuoka-ken.ac.jp
}
%%%%%%%
\begin{abstract}
Casimir energy is calculated for the 5D electromagnetism 
in the {\it warped} geometry. 
It is compared with the {\it flat} case(arXiv:0801.3064). 
%The position/momentum propagator is exploited. 
A new regularization, 
called {\it sphere lattice regularization}, is taken. 
It is based on the {\it minimal area principle} and 
is a {\it direct} realization of the geometrical approach 
to the {\it renormalization group}. 
The properly regularized form of Casimir energy, is expressed in a closed form. 
We numerically evaluate $\La$(4D UV-cutoff), $\om$(5D bulk curvature, 
warp parameter)
and $T$(extra space IR parameter) dependence of the Casimir energy. 
The {\it warp parameter} $\om$ suffers from the {\it renormalization effect}. 
We examine the meaning of the weight function and finally 
reach a {\it new definition} of the Casimir energy where {\it the 4D momenta( or coordinates) 
are quantized} with the extra coordinate as the Euclidean time. We comment on the cosmological term at the end. 
\end{abstract}
\keywords{
%warped geometry; position/momentum propagator; heat kernel; Casimir energy;  
sphere lattice; renormalization of boundary parameters.
%cosmological constant.
}

\bodymatter
%%%%
%%%%%%%%%%%%%%%%%%%%%%%%%%%%%%%%%%%%%%%%%%%%%%%%%%%%%%%%%%%%%%%%%%%%%%%%%
%%%%%%%%%%%%%%%%%%%%%%%%%%%  Sec.1   %%%%%%%%%%%%%%%%%%%%%%%%%%%%%%%%%%%%
\section
{Introduction}\label{intro}
%***label**{intro}
%%%%%%%%%%%%%%%%%%%%%%%%%%%%%%%%%%%%%%%%%%%%%%%%%%%%%%%%%%%%%%%%%%%%%%%%%
%%%%%%%%%%%%%%%%%%%%%%%%%%%%%%%%%%%%%%%%%%%%%%%%%%%%%%%%%%%%%%%%%%%%%%%%%
In the previous work\cite{SI0801}, we have examined 5D electromagnetism
in the {\it flat} geometry. 
The extra space ($y$) is {\it periodic} (periodicity $2l$) and
$Z_2$-symmetry is taken into account.  
The (regularized) Casimir energy is expressed in a closed form:\ 
%*** UIreg2 %%%%%%%%%%%%%%%%
%\bea
$
E_{Cas}(\La,l)=\frac{2\pi^2}{(2\pi)^4}\int_{1/l}^{\La}d\ptil\int_{1/\La}^ldy~\ptil^3 
W(\ptil,y)F(\ptil,y),\ 
F(\ptil,y)\equiv 
\int_\ptil^\La d\ktil\frac{-3\cosh\ktil(2y-l)-5\cosh\ktil l}{2\sinh(\ktil l)}
,\ 
%\label{UIreg2}
$
%\eea
%%%%%%%%%%%%%%%%%%%%%%%%%%%%%
where 
$\La$ is the 4D-momentum cutoff, $W(\ptil,y)$ is a {\it weight function} 
to suppress the IR and UV divergences. The past approach ($W=1$) tells us 
the quintic divergence ($\La^5$) of the energy. This fact has been troubling 
us as the problem of the divergent cosmological constant in the 5D Kaluza-Klein theory\cite{Kal21,Klein26,AC83}.
%
%%%%%%%%%%%%%%%%%%%%%%%%%%%%  Sec.2  %%%%%%%%%%%%%%%%%%%%%%%%%%%%%%%%%
%%%%   Casimir Energy of 4D Electromagnetism     %%%%%%%%%%%
%%%%%%%%%%%%%%%%%%%%%%%%%%%%%%%%%%%%%%%%%%%%%%%%%%%%%%%%%%%%%%%%%%%%%%
%\section
%{Casimir Energy of 4D Electromagnetism}\label{4dEM}
%***label**{4dEM}

Let us review Casimir energy in 4D Electromagnetism. 
The electromagnetic field within two parallelly-placed (along z-direction, 
separation length $2l$) perfectly-conducting plates can be regarded 
as the sum of harmonic osscilators. 
For the x,y-directions, we take the periodic (periodicity $2L$) regularization.  
The energy of the 4D EM is given by
%*** 4dEM18 %%%%%%%%%%%%%%%%
\bea
E_{4dEM}=E_{Cas}+E_\be\ ,\ 
E_{Cas}=\sum \omtil_{m_xm_yn}\ ,\ 
E_{\be}=2\sum \frac{\omtil_{m_xm_yn}}{\e^{\be\omtil_{m_xm_yn}}-1}
\ ,
\label{4dEM18}
\eea 
%%%%%%%%%%%%%%%%%%%%%%%%%%%
where ${\omtil_{m_xm_yn}}^2=(\omvec_{m_xm_yn})^2=(m_x\frac{\pi}{L})^2+(m_y\frac{\pi}{L})^2+(n\frac{\pi}{l})^2$. 

\begin{figure}
\begin{center}
\parbox{60mm}{
\epsfig{figure=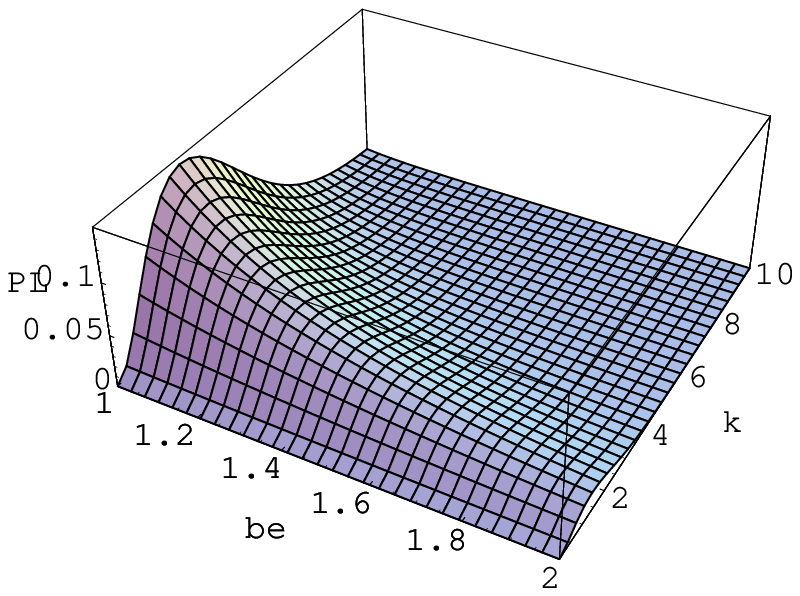,height=4cm}
%\figsubcap{a}
               }
                            %%%      %%%
\hspace*{2mm}
                            %%%      %%%
\parbox{60mm}{
\epsfig{figure=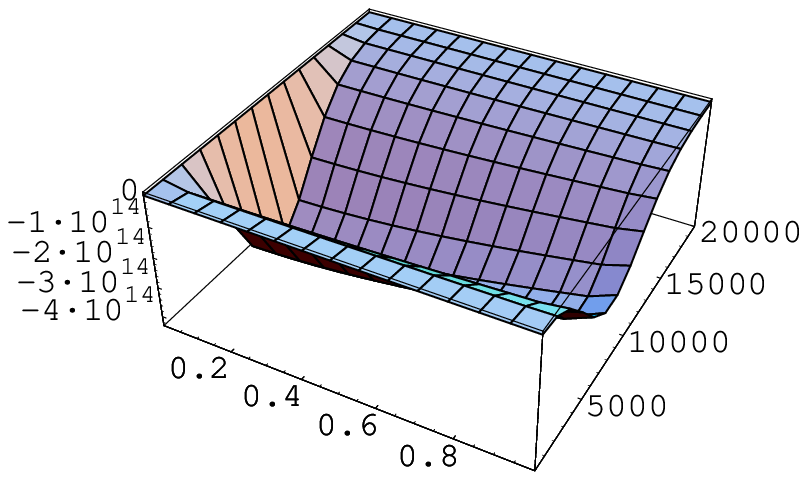,height=4cm}
%\figsubcap{b}
               }
\caption{(a-left) Graph of Planck's radiation formula.  
$ \Pcal (\be,k)=\frac{1}{(c\hbar)^3}\frac{1}{\pi^2}k^3/(\e^{\be k}-1)\ \ 
(1\leq\be\leq 2,\ 0.01\leq k\leq 10)$.  
%***PlanckDistB.eps\
\ (b-right) Behavior of $(-N_1/2)\ptil^3W_1(\ptil,z)F^-(\ptil,z)$(elliptic suppression). 
$\La=20000,\ \om=5000,\ T=1$\ . 
$1.0001/\om\leq z\leq 0.9999/T ,\ \m=\La T/\om\leq \ptil\leq \La$. 
%***W1L2mank5senT1.eps
        }
\label{PlanckDistBW1L2mank5senT1}
\end{center}
\end{figure}

$E_\be$ gives us Stefan-Boltzmann's law. 
%*** 4dEM19 %%%%%%%%%%%%%%%%
\bea
E_{\be}
=2\int_0^\infty\frac{dk4\pi k^2}{(\frac{\pi}{L})^2\frac{\pi}{l}}\frac{\omtil(k)}{\e^{\be\omtil(k)}-1}
=(2L)^2(2l)\int_0^\infty dk \Pcal(\be,k) 
=\frac{8}{\pi^2}\frac{L^2l}{\be^4}3! \zeta(4)
\com
\label{4dEM19}
\eea 
%%%%%%%%%%%%%%%%%%%%%%%%%%%
where $\omtil(k)=k$. 
$\Pcal(\be,k)$ is the Planck's radiation formula. The behavior of $\Pcal(\be,k)$ is 
graphically shown in Fig.\ref{PlanckDistBW1L2mank5senT1}a. (We will see similar graphs in the following 
5D case. The extra axis corresponds to the $\be$-axis. ) 
The peak curve of the graph is {\it hyperbolic}, $\be k$=const., in the $(\be,k)$-plane 
(Wien's displacement law). 
                             %%%   <Fig.   %%%
                              %%%%   Fig.**.>    %%%%%

$E_{Cas}$ is the sum of the {\it zero-point energy} over the all 
frequencies. It is 
{\it Casimir energy}. This quantity is {\it independent
of the coupling} and is {\it dependent on the boundary(ies)}.   
$E_{Cas}$ does not vanish for the temperature($1/\be$)=0. 
It is, however, formally {\it divergent}.  We need a proper {\it regularization} 
for the {\it summation over the infinite degree of freedom} due to the continuity 
of the space-time.\q $E^\La_{Cas}=$
%*** 4dEM20 %%%%%%%%%%%%%%%%
\bea
\sum\omtil_{m_xm_yn}~g(\frac{\omtil_{m_xm_yn}}{\La})
=\sum\sqrt{
(m_x\frac{\pi}{L})^2+(m_y\frac{\pi}{L})^2+(n\frac{\pi}{l})^2
                              }~g(\frac{\omtil_{m_xm_yn}}{\La})
,
\label{4dEM20}
\eea 
%%%%%%%%%%%%%%%%%%%%%%%%%%%
where we introduce the {\it cut-off} function: 
$g(\om/\La)=\{~1 \mbox{\ for\ } 0\leq\om\leq\La;\ 0 \mbox{\ for\ } \om>\La\}$. 
$\La$ is the 
cut-off parameter for the absolute value of the 3D (x,y,z) momentum. 
We will take the limit $\La\ra\infty$ at an appropriate stage. 
Taking the reference point, $L\ra\infty, L\gg l\ra\infty$, from which we "measure" the energy, 
%*** 4dEM21 %%%%%%%%%%%%%%%%
%\bea
$
E^{\La 0}_{Cas}=
\int_{-\infty}^\infty\int_{-\infty}^\infty\frac{dk_xdk_y}{(\pi/L)^2} 
\int_{-\infty}^\infty\frac{dk_z}{\pi/l} 
\sqrt{k_x^2+k_y^2+k_z^2}~g(\frac{k}{\La})
\ ,
$
%\label{4dEM21}
%\eea 
%%%%%%%%%%%%%%%%%%%%%%%%%%%
we finally obtain the {\it finite} result, 
%*** 4dEM24 %%%%%%%%%%%%%%%%
%\bea
$
u=\frac{E^\La_{Cas}-E^{\La 0}_{Cas}}{(2L)^2}=\frac{\pi^2}{(2l)^3}\frac{B_4}{4!}=-\frac{\pi^2}{720}\frac{1}{(2l)^3}
\ ,
$
%\label{4dEM24}
%\eea 
%%%%%%%%%%%%%%%%%%%%%%%%%%%
which deoes not depend on $\La$. Especially there remains no log$\La$ divergence. 
This point is contrasting with the ordinary renormalization of {\it interacting} theories 
such as 4D QED and 4D YM. 
Hence we need {\it not} the renormalization of the wave-function and the parameter $l$. 
In the 5D flat geometry, 
the {\it renormalization of the boundary parameter} is necessary\cite{SI0801}. 
In the following, we examine the warped 5D case and find the another boundary parameter $\om$ 
suffers from the renormalization effect.

%%%%%%%%%%%%%%%%%%%%%%%%%%%%  Sec.2  %%%%%%%%%%%%%%%%%%%%%%%%%%%%%%%%%
%%%%   Kaluza-Klein expansion approach     %%%%%%%%%%%
%%%%%%%%%%%%%%%%%%%%%%%%%%%%%%%%%%%%%%%%%%%%%%%%%%%%%%%%%%%%%%%%%%%%%%
\section{
{Kaluza-Klein expansion approach\label{KKexp}}
}
%***label***{KKexp}
5D {\it massive} vector theory is given by 
%*** KKexp1%%%%%%%%%%%%%%%%
\bea
S_{5dV}=\intxz\sqrt{-G}(-\fourth F_{MN}F^{MN}-\half m^2A^M A_M),
F_{MN}=\pl_M A_N-\pl_NA_M,\nn
ds^2=\frac{1}{\om^2z^2}(\eta_\mn dx^\m dx^\n+{dz}^2)=G_{MN}dX^M dX^N,
G\equiv\det G_{AB},
\label{KKexp1}
\eea 
%%%%%%%%%%%%%%%%%%%%%%%%%%%
where 
$
M,N=0,1,2,3,5(\mbox{or }z);\ \mu,\nu=0,1,2,3
$. 
Casimir energy is given by some integral where the (modifed) Bessel functions, 
with the index $\nu=\sqrt{1+\frac{m^2}{\om^2}}$, appear. 
Hence the 5D EM 
limit is given by $\n=1\ (m=0)$. We consider, however, the {\it imaginary} 
mass case $m=i\om\ (m^2=-\om^2,\ \n=0)$. 
Instead of analyzing the $m^2=-\om^2$ of the massive vector (\ref{KKexp1}), 
we take the 5D massive {\it scalar} theory
on AdS$_5$ with $m^2=-4\om^2,\ \n=\sqrt{4+m^2/\om^2}=0$. 
%*** KKexp1b%%%%%%%%%%%%%%%%
\bea
\Lcal=\sqrt{-G}(-\half \na^A\Phi\na_A\Phi-\half m^2\Phi^2)\com\q 
\na^A\na_A \Phi-m^2\Phi+J=0\com 
\label{KKexp1b}
\eea 
%%%%%%%%%%%%%%%%%%%%%%%%%%%
where 
$G\equiv \det G_{AB}\ ,\ 
ds^2=G_{AB}dX^AdX^B\ $. 
$\Phi(X)=\Phi(x^a,z)$ is the 5D scalar field. 
The background geometry is AdS$_5$ which takes the form:\ 
%*** KKexp2 %%%%%%%%%%%%%%%%
%\bea
$
(G_{AB})=diag(
\frac{1}{\om^2z^2}\eta_{ab},\frac{1}{\om^2z^2}
              )
,\ 
\sqrt{-G}=\frac{1}{(\om |z|)^5}.
$\ 
The variable range of $z$ is 
$
-\frac{1}{T}\leq z \leq -\frac{1}{\om}\q\mbox{or}\q
\frac{1}{\om}\leq z \leq \frac{1}{T}\q 
(-l\leq y \leq l\ ,\ |z|=\frac{1}{\om}\e^{\om |y|}), 
$
\ where $l$ and $T$ are related as 
$
\frac{1}{T}=\frac{1}{\om}\e^{\om l}.
$
%\label{KKexp2}
%\eea 
%%%%%%%%%%%%%%%%%%%%%%%%%%%
We take into account $Z_2$ symmetry: $z\change -z$.  
$\om$ is the bulk curvature (AdS$_5$ parameter) and $T^{-1}$ is the
size of the extra space (infrared regularization parameter). 

The Casimir energy $E_{Cas}$ is given by\ 
$
\e^{-T^{-4}E_{Cas}}=\int\Dcal\Phi\exp\{i\int d^5X\Lcal\}=
$
%*** KKexp3 %%%%%%%%%%%%%%%%
\bea
\int\Dcal\Phi(X)\exp\left[
i\intfx dz\frac{1}{(\om|z|)^5}\half\Phi\{
\om^2z^2\pl_a\pl^a\Phi+(\om |z|)^5\frac{\pl}{\pl z}\frac{1}{(\om z)^3}\pl_z\Phi
                           -m^2\Phi
                                       \} 
                     \right]
\ .
\label{KKexp3}
\eea 
%%%%%%%%%%%%%%%%%%%%%%%%%%%
Here we introduce, instead of $\Phi(X)$, the partially (4D world only)
Fourier transformed field $\Phi_p(z)$:\  
%*** KKexp4 %%%%%%%%%%%%%%%%
%\bea
$
\Phi(X)=\intp \e^{ipx}\Phi_p(z)\ .
$
%\label{KKexp4}
%\eea 
%%%%%%%%%%%%%%%%%%%%%%%%%%% 
Eq.(\ref{KKexp3}) can be rewritten as
%*** KKexp5 %%%%%%%%%%%%%%%%
\bea
s(z)\equiv\frac{1}{(\om z)^3}\com\q
\Lhat_z\equiv\frac{d}{dz}\frac{1}{(\om z)^3}\frac{d}{dz}
                           -\frac{m^2}{(\om z)^5}
\com\q
\e^{-T^{-4}E_{Cas}}
=\int\Dcal\Phi_p(z)\times\nn
\exp\left[
i\intp 2\int_{1/\om}^{1/T}dz\left\{
\half\Phi_p(z)\{
-\frac{1}{(\om z)^3}p^2+\frac{d}{dz}\frac{1}{(\om z)^3}\frac{d}{dz}
                           -\frac{m^2}{(\om z)^5}
                \}\Phi_p(z)
                             \right\} 
                          \right].
\label{KKexp5}
\eea 
%%%%%%%%%%%%%%%%%%%%%%%%%%%
%*** KKexp6 %%%%%%%%%%%%%%%%
%\bea
%
%\label{KKexp6}
%\eea 
%%%%%%%%%%%%%%%%%%%%%%%%%%%
We consider the Bessel eigen-value problem:\ 
$
\{s(z)^{-1}\Lhat_z+{M_n}^2\}\psi_n(z)=0,\ 
$
with $Z_2$-property:\ 
%*** KKexp7 %%%%%%%%%%%%%%%%
%\bea
$
\psi_n(z)=-\psi_n(-z)\ \mbox{for}\ P=-\ ;\ 
\psi_n(z)=\psi_n(-z)\ \mbox{for}\ P=+ 
\ ,\ 
$
%\label{KKexp7}
%\eea
%%%%%%%%%%%%%%%%%%%%%%%%%%%
and the appropriate b.c. at fixed points. Because the set $\{\psi_n(z)\}$ 
constitute the orthonormal and complete system, we can express $\Phi_p(z)$ as, 
%*** KKexp8 %%%%%%%%%%%%%%%%
%\bea
$
\Phi_p(z)=\sum_{n}c_n(p)\psi_n(z)
\ .
$
%\label{KKexp8}
%\eea
%%%%%%%%%%%%%%%%%%%%%%%%%%%
\ Eq.(\ref{KKexp5}) can be further rewritten as\ \ $\e^{-T^{-4}E_{Cas}}=$ 
%*** KKexp9 %%%%%%%%%%%%%%%%
\bea
\int\Dcal\Phi_p(z)\exp\left[
i\intp 2\int_{1/\om}^{1/T}dz\left\{
\half\Phi_p(z)s(z)( {s(z)}^{-1}\Lhat_z-p^2)\Phi_p(z)
                             \right\} 
                          \right]\nn
=\int\prod_n dc_n(p)\exp\left[
i\intp\sum_n\{
\frac{-1}{2} c_n(p)^2(p^2+M_n^2)
             \}
                         \right]
=\exp\sum_{n,p}\{\frac{-1}{2}
\ln (p^2+M_n^2)\},
\label{KKexp9}
\eea 
%%%%%%%%%%%%%%%%%%%%%%%%%%%
where the orthonormal relation,\ 
%*** KKexp10 %%%%%%%%%%%%%%%%
%\bea
$
2\int_{\frac{1}{\om}}^{\frac{1}{T}}\psi_n(z)s(z)\psi_m(z)dz
=\del_{nm}
\ ,
$
%\label{KKexp10}
%\eea 
%%%%%%%%%%%%%%%%%%%%%%%%%%%
\ is used. 
This shows that $s(z)$, defined in (\ref{KKexp5}), plays the role of 
"inner product measure" in the function space 
$\{\psi_n(z), 1/\om \leq z\leq 1/T\}$. 
The expression (\ref{KKexp9}) is the familiar one (expanded form) 
of the Casimir energy.

%%%%%%%%%%%%%%%%%%%%%%%%%%%%  Sec.3  %%%%%%%%%%%%%%%%%%%%%%%%%%%%%%%%%
%%%%   Heat-Kernel Approach and Position/Momentum Propagator    %%%%%%
%%%%%%%%%%%%%%%%%%%%%%%%%%%%%%%%%%%%%%%%%%%%%%%%%%%%%%%%%%%%%%%%%%%%%%
\section{
{Heat-Kernel Approach and Position/Momentum Propagator\label{HKA}}
}
%***label***{HKA}
Instead of the KK-expansion form, we can formally integrate out the
$\Phi_p(z)$ variable in the path-integral (\ref{KKexp9}). 
%*** HKA1 %%%%%%%%%%%%%%%%
\bea
\e^{-T^{-4}E_{Cas}}
=\exp\left[
T^{-3}\intp 2\int_{1/\om}^{1/T}dz s(z)\left\{
-\half\ln ( -{s(z)}^{-1}\Lhat_z+p^2)
                             \right\} 
      \right]                      \nn
=\exp\left[
T^{-3}\intp 2\int_{1/\om}^{1/T}dz s(z)\left\{
\half\int_0^\infty\frac{1}{t}\e^{ t( {s(z)}^{-1}\Lhat_z-p^2)}dt +\mbox{const}
                            \right\} 
      \right]                      
,
\label{HKA1}
\eea 
%%%%%%%%%%%%%%%%%%%%%%%%%%%
The above formal result can be {\it precisely} defined using the heat 
equation\cite{Schwinger51}. 
%*** HKA2%%%%%%%%%%%%%%%%
\bea
\e^{-T^{-4} E_{Cas}}
=(\mbox{const})\times\exp T^{-4}\intp 2\int_{0}^{\infty}\half\frac{dt}{t}
\mbox{Tr}~H_p(z,z';t)\com\nn
\mbox{Tr}~H_p(z,z';t)=\int_{1/\om}^{1/T}s(z)H_p(z,z;t)dz,
\{\frac{\pl}{\pl t}-(s^{-1}\Lhat_z-p^2) \}H_p(z,z';t)=0
.
\label{HKA2}
\eea
%%%%%%%%%%%%%%%%%%%%%%%%%%%%%
The heat kernel $H_p(z,z';t)$ is formally solved, using the
Dirac's bra and ket vectors $(z|, |z)$, as\ 
%*** HKA3%%%%%%%%%%%%%%%%
%\bea
$
H_p(z,z';t)=(z|\e^{-(-s^{-1}\Lhat_z+p^2)t}|z')
\ .
$
%\label{HKA3}
%\eea
%%%%%%%%%%%%%%%%%%%%%%%%%%%%%
\ Using the set $\{\psi_n(z)\}$ defined previously, the explicit solution of (\ref{HKA2}) is
given by
%*** HKA4%%%%%%%%%%%%%%%%
\bea
\left\{
\begin{array}{c}
H_p(z,z';t)\\
E_p(z,z';t)
\end{array}
\right\}
=\sum_{n\in \bfZ}\e^{-(M_n^2+p^2)t}
\half\{ \psi_n(z)\psi_n(z')\mp\psi_n(z)\psi_n(-z') \},\ P=\mp
\ ,
\label{HKA4}
\eea
%%%%%%%%%%%%%%%%%%%%%%%%%%%%%
We here introduce the position/momentum propagators $G^{\mp}_p$ as
follows.\ \ $G^\mp_p(z,z')\equiv$
%*** HKA6%%%%%%%%%%%%%%%%
\bea
\int_0^\infty dt \left\{
\begin{array}{c}
H_p(z,z';t)\\
E_p(z,z';t)
\end{array}
                  \right\}
=
\sum_{n\in \bfZ}\frac{1}{M_n^2+p^2}
\half\{ \psi_n(z)\psi_n(z')\mp\psi_n(z)\psi_n(-z') \}
.
\label{HKA6}
\eea
%%%%%%%%%%%%%%%%%%%%%%%%%%%%%
Therefore the Casimir energy $E_{Cas}$ is given by
%*** HKA8%%%%%%%%%%%%%%%%
\bea
E_{Cas}(\om,T)=
\intp 2\intt 2\int_{1/\om}^{1/T} dz~s(z)H_p(z,z;t)\nn
=\intp 2\intt 2\int_{1/\om}^{1/T} dz~s(z)\left\{
\sum_{n\in \bfZ}\e^{-(M_n^2+p^2)t}\psi_n(z)^2
                                       \right\}
\com
\label{HKA8}
\eea
%%%%%%%%%%%%%%%%%%%%%%%%%%%%%
%This expression leads to the same treatment as the previous section. 

\begin{figure}\begin{center}
\parbox{60mm}{
\epsfig{figure=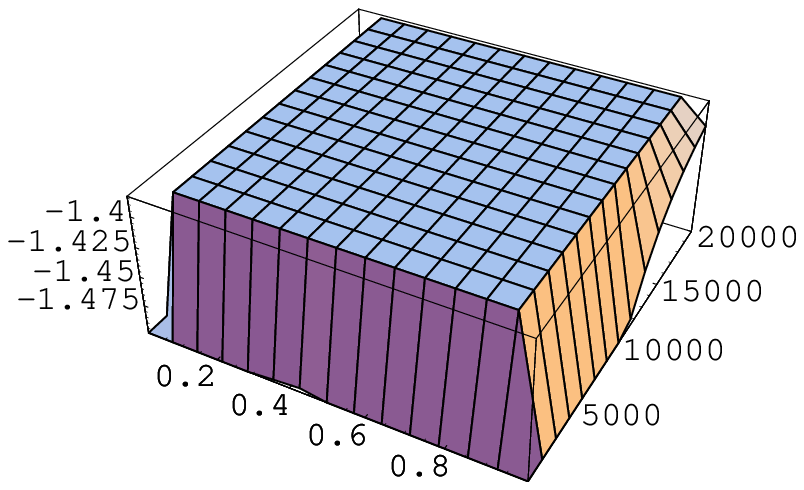,height=4cm}
              }
                      %%%%      %%%%%
\hspace*{2mm}
                      %%%%%     %%%%%
\parbox{60mm}{
\epsfig{figure=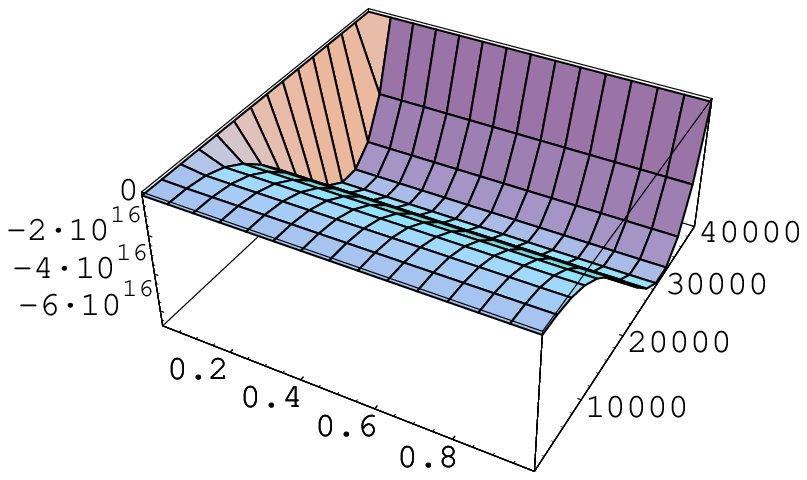,height=4cm}
              }
\caption{(a-left)\ 
Behavior of $\ln |\half\Fcal^-(\ktil,z)|=\ln |\ktil~ G^-_k(z,z)/(\om z)^3|$. 
$\om=10^4, T=1, \La=2\times 10^4$. $1.0001/\om \leq z \leq 0.9999/T$. $\La T/\om \leq \ktil \leq \La$. 
Note $\ln |(1/2)\times (-1/2)|\approx -1.39$.  
%***FcalmHT1k10p4.eps
\ (b-right)\ 
Behaviour of $(-1/2)\ptil^3F^-(\ptil,z)$ (\ref{UIreg2b}). $T=1, \om=10^4, \La=4\cdot10^4$.  
$1.0001/\om\leq z<0.9999/T$, $\La T/\om\leq\ptil\leq \La$ . 
%***p3FmL40000.eps
        }
\label{FcalmHT1k10p4p3FmL40000}
\end{center}\end{figure}

The P/M propagators $G_p^\mp$ in (\ref{HKA6}) can be expressed in a {\it closed} form.
(See, for example, Ref.~\refcite{IM0703}.) %***time-like no baai mo kakuka?*** 
Taking the {\it Dirichlet} condition at all fixed points, the expression
for the fundamental region ($1/\om \leq z\leq z'\leq 1/T$) is given by
\ \ $G_p^\mp(z,z')=$
%*** HKA12 %%%%%%%%%%%%%%%%
\bea
\mp\frac{\om^3}{2}z^2{z'}^2
\frac{\{\I_0(\Pla)\K_0(\ptil z)\mp\K_0(\Pla)\I_0(\ptil z)\}  
      \{\I_0(\Tev)\K_0(\ptil z')\mp\K_0(\Tev)\I_0(\ptil z')\}
     }{\I_0(\Tev)\K_0(\Pla)-\K_0(\Tev)\I_0(\Pla)}
\ ,
\label{HKA12}
\eea 
%%%%%%%%%%%%%%%%%%%%%%%%%%% 
where $\ptil\equiv\sqrt{p^2}\com\q p^2\geq 0\ (\mbox{space-like})$. 
We can express the $\La$-regularized Casimir energy in terms of the following 
functions $F^\mp(\ptil,z)$. 
%*** HKA13%%%%%%%%%%%%%%%%
\bea
E^{\La,\mp}_{Cas}(\om,T)=
\left.\intp\right|_{\ptil\leq\La}\int_{1/\om}^{1/T}dz~F^\mp(\ptil,z)\com
\label{HKA13}
\eea
%%%%%%%%%%%%%%%%%%%%%%%%%%%%% 
where 
$
F^\mp(\ptil,z)\equiv s(z)\int_{p^2}^{\La^2}\{G_k^\mp(z,z) \}dk^2
\equiv\int_\ptil^\La\Fcal^\mp(\ktil,z)d\ktil
$. 
Here we introduce the UV cut-off parameter $\La$ for the 4D momentum space. 
In Fig.\ref{FcalmHT1k10p4p3FmL40000}a, we show the behaviour of $\Fcal^-(\ktil,z)$. 
The table-shape graph says the "Rayley-Jeans" dominance.
%{\it equal-weight distribution} of degree of freedom over $(\ktil,z)$-space. 
That is, for the wide-range region $(\ptil,z)$ satisfying both 
$\ptil (z-\frac{1}{\om})\gg 1$ and $\ptil (\frac{1}{T}-z)\gg 1$, 
%*** HKA14%%%%%%%%%%%%%%%%
\bea
\Fcal^-(\ptil,z)\approx \frac{-1}{2},
\Fcal^+(\ptil,z)\approx \frac{-1}{2};\ 
\ptil (z-\frac{1}{\om})\gg 1\ \mbox{and}\ \ptil (\frac{1}{T}-z)\gg 1.
\label{HKA14}
\eea
%%%%%%%%%%%%%%%%%%%%%%%%%%%%%
                             %%%   <Fig.1   %%%
                              %%%   Fig.***>  %%%

%%%%%%%%%%%%%%%%%%%%%%%%%%%%  Sec.4  %%%%%%%%%%%%%%%%%%%%%%%%%%%%%%%%% 
%%%% UV and IR Regularization Parameters and Evaluation   %%%
%%%%%%%%%%%%%%%%%%%%%%%%%%%%%%%%%%%%%%%%%%%%%%%%%%%%%%%%%%%%%%%%%%%%%%
\section{
{UV and IR Regularization Parameters and Evaluation of Casimir Energy\label{UIreg}}
}
%***label***{UIreg}

\begin{figure}\begin{center}
\parbox{60mm}{
\epsfig{figure=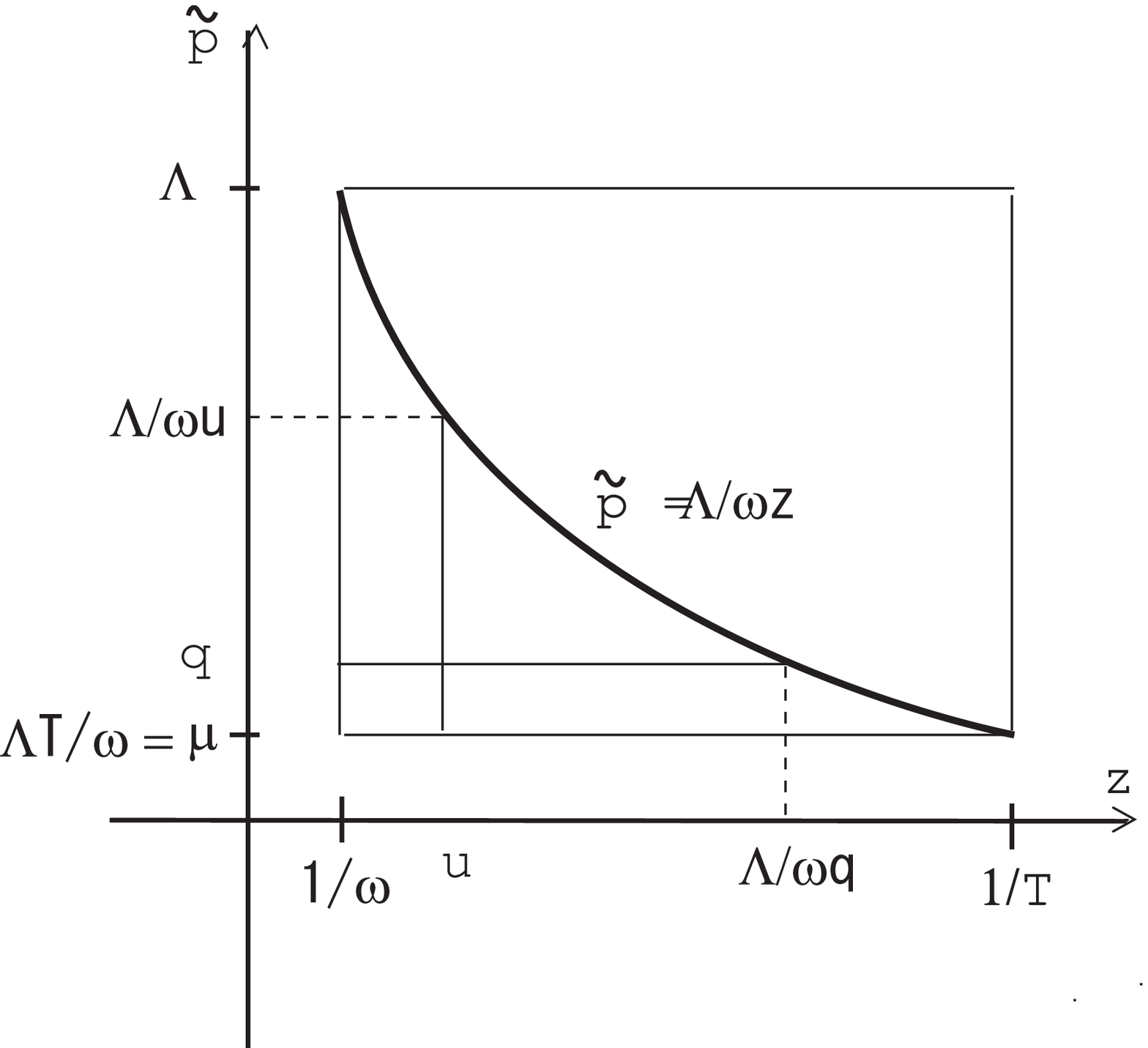,height=4cm}
              }
              \hspace*{2mm}
\parbox{60mm}{
\epsfig{figure=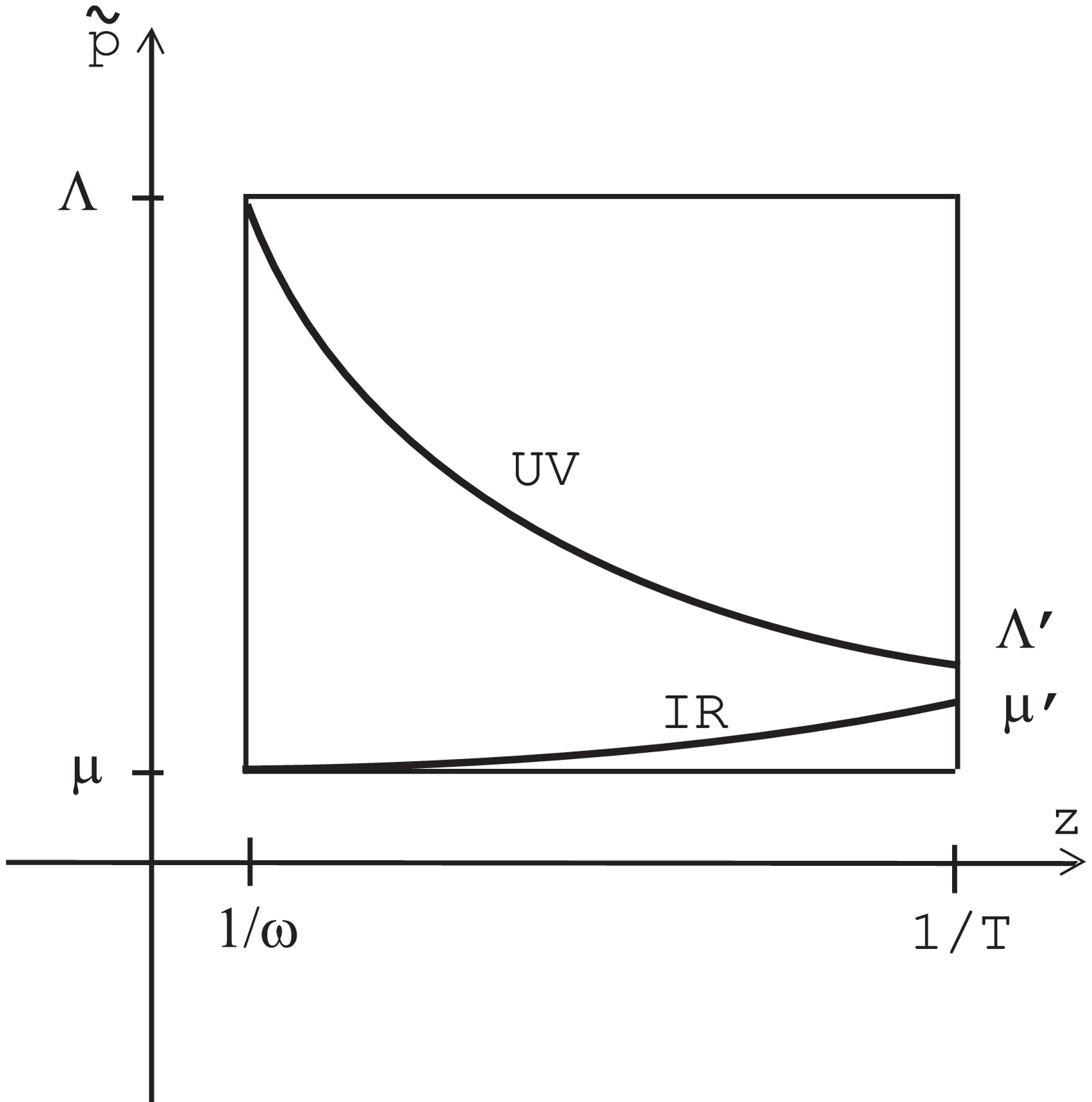,height=4cm}
              }
\caption{(a-left)\ 
Space of (z,$\ptil$) for the integration. The hyperbolic curve 
will be used in Sec.\ref{surf}.  
%***zpINTregionW.eps
\ (b-right)\ 
Space of ($\ptil$,z) for the integration (present proposal). 
%***zpINTregionW2.eps
}
\label{zpINTregionWzpINTregionW2}
\end{center}\end{figure}

The integral region of the above equation (\ref{HKA13}) is displayed in Fig.\ref{zpINTregionWzpINTregionW2}a. 
In the figure, we introduce the regularization cut-offs for the 4D-momentum integral, 
$\m\leq\ptil\leq\La$. 
As for the extra-coordinate integral, it is the finite interval, 
$1/\om\leq z\leq 1/T=\e^{\om l}/\om$, hence we need not introduce further
regularization parameters. 
%In order to supress the number of artificial parameters as much as possible
For simplicity, we take
the following IR cutoff of 4D momentum
:\ 
%*** HKA15%%%%%%%%%%%%%%%%
%\bea
$
\m=\La\cdot\frac{T}{\om}=\La \e^{-\om l}
\ .
$
%\label{HKA15}
%\eea
%%%%%%%%%%%%%%%%%%%%%%%%%%%%%
\ 
%Hence the new regularization parameter is $\La$ only. 
%This is the same situation as the lattice gauge theory. 
%(the unit lattice size = $1/\om$, the total lattice size =$1/T$.)
                             %%%   <Fig.3   %%%
                              %%%   Fig.***>  %%%

Let us evaluate the ($\La,T$)-{\it regularized} value of (\ref{HKA13}). 
%*** UIreg2b %%%%%%%%%%%%%%%%
\bea
E_{Cas}^{\La,\mp}(\om,T)=\frac{2\pi^2}{(2\pi)^4}\int_{\m}^{\La}d\ptil\int_{1/\om}^{1/T}dz~\ptil^3 F^\mp (\ptil,z)\com
\label{UIreg2b}
\eea
%%%%%%%%%%%%%%%%%%%%%%%%%%%%%
where
$
F^\mp (\ptil,z)
=\frac{2}{(\om z)^3}\int_\ptil^\La\ktil~ G^\mp_k(z,z)d\ktil
$. 
The integral region of ($\ptil,z$) is the {\it rectangle} shown in Fig.\ref{zpINTregionWzpINTregionW2}a
.

Note that eq.(\ref{UIreg2b}) is the {\it rigorous} expression of the $(\La,T)$-regularized Casimir energy. 
We show the behaviour of $(-1/2)\ptil^3F^-(\ptil,z)$ in Fig.\ref{FcalmHT1k10p4p3FmL40000}b. 
%footnote
\ [
\small{
The requirement for the three parameters $\om, T, \La$ is $\La\gg\om\gg T$. 
}
]\ 
From a {\it close} numerical analysis of ($\ptil,z$)-integral (\ref{UIreg2b})
, 
we have confirmed:\ 
%*** UIreg5 %%%%%%%%%%%%%%%%
%\bea
$
E^{\La,-}_{Cas}(\om,T)=\frac{2\pi^2}{(2\pi)^4}\times\left[ -0.0250 \frac{\La^5}{T} \right]
\ ,
$
%\label{UIreg5}
%\eea
%%%%%%%%%%%%%%%%%%%%%%%%%%%%%
which does {\it not} depend on $\om$. 
%$E^{\La,+}_{Cas}(\om,T)$\ MO YARUKA ??? 
%** CasEneWarpSca071126.nb,  L-dependence $T$-dependence Mathematica Cal(Scalar -+) ****
%(Note: $0.025=1/40$.) 
$\ln\frac{\La}{T}$-term does not appear. 
Compared with the flat case 
($
E_{Cas}^{flat}(\La,l)=(1/8\pi^2)\left[ 
-0.1249 l\La^5
-(1.41, 0.706, 0.353)\times 10^{-5}~l\La^5\ln (l\La)
                               \right] 
$
), 
we see the factor $T^{-1}$ plays the role of {\it IR cut-off} of the
extra space. We note that the behavior of Fig.\ref{FcalmHT1k10p4p3FmL40000}b 
is similar to the Rayleigh-Jeans's region (small momentum region) of the Planck's radiation 
formula (Fig.\ref{PlanckDistBW1L2mank5senT1}a) in the sense that 
$\ptil^3F(\ptil,z)\propto \ptil^3$ for $\ptil\ll\La$. 

From the Fig.\ref{FcalmHT1k10p4p3FmL40000}b, the approximate form 
of $F(\ptil,z)$ for the large $\La$ and $1/T$ is given by:\ 
%*** UIreg6 %%%%%%%%%%%%%%%%
%\bea
$
F^\mp (\ptil,z)\approx -\frac{f}{2} \La (1-\frac{\ptil}{\La})\ ; f=1
\ ,
$
%\label{UIreg6}
%\eea
%%%%%%%%%%%%%%%%%%%%%%%%%%%%%
\ which does {\it not} depend on $z, \om$ and $T$. $f$ is the degree of freedom.

%%%%%%%%%%%%%%%%%%%%%%%%%%%%  Sec.6  %%%%%%%%%%%%%%%%%%%%%%%%%%%%%%%%%
%%%% UV and IR Regularization Surfaces and Principle of Minimal Area  %%%
%%%%%%%%%%%%%%%%%%%%%%%%%%%%%%%%%%%%%%%%%%%%%%%%%%%%%%%%%%%%%%%%%%%%%%
\section{
UV and IR Regularization Surfaces, Principle of 
Minimal Area and Renormalization Flow\label{surf}
}
%***label***{surf}
In order to avoid the $\La^5$-divergence, 
a proposal was presented by Randall and Schwartz\cite{RS01}. They introduced
the {\it position-dependent cut-off},\ $\mu <\ptil <\La /\om u\ ,\ u\in [1/\om,1/T]$\ , 
for the 4D-momentum integral in the "brane" located at $z=u$. See Fig.\ref{zpINTregionWzpINTregionW2}a.
We have confirmed that the value $E_{Cas}$ of (\ref{UIreg2b}), when the Randall-Schwartz 
integral region (Fig.\ref{zpINTregionWzpINTregionW2}a) is taken, is proportional to $\La^5$:\ 
%*** surfM1 %%%%%%%%%%%%%%%%
%\bea
$
E^{-RS}_{Cas}(\om,T)=
\frac{2\pi^2}{(2\pi)^4}\frac{\La^5}{\om}\left\{
-1.58\times 10^{-2}-1.69\times 10^{-4}\ln~\frac{\La}{\om}
                                          \right\}
\ ,
$
%\label{surfM1}
%\eea
%%%%%%%%%%%%%%%%%%%%%%%%%%%%%
%**LeastSqFit070924.nb*** 
which is {\it independent} of $T$ 
%(within the described significant digits)
.
This shows the divergence 
situation does {\it not} improve compared with the non-restricted case 
: $E^{\La,-}_{Cas}$ in Sec.4. 
%of (\ref{UIreg5})
. 
{\it $T$ of $E^{\La,-}_{Cas}$ is replaced by the warp parameter $\om$.} 
These are contrasting with the flat case 
($
E^{RS,flat}_{Cas}=(1/8\pi^2)[-0.0893~\La^4]
$;\ No $\ln(l\La)$-term
)
. 

Although they claim the holography is behind the procedure, 
the legitimateness of the restriction looks less obvious. We have proposed 
an alternate approach 
and given a legitimate explanation within the 5D QFT\cite{IM0703,SI07Nara,SI0801,SI0803OCU}. 
See Fig.~\ref{zpINTregionWzpINTregionW2}b. 
The restriction region is bounded by two minimal surfaces, IR-surface and 
UV-surface. They are 4 dimensional manifold made from $S^3$-spheres "running" 
along $z$-axis. On the 'brane' at fixed $z$, 
the 4D region bounded by IR-surface is regarded as a large-size 4D-ball and 
that by UV-surface is regarded as a small-size 4D-ball. Hence 
this regularization configuration is the 4D {\it sphere lattice} running 
along $z$-axis. See Ref.~\refcite{SI0812}.

%%%%%%%%%%%%%%%%%%%%%%%%%%%%  Sec.6  %%%%%%%%%%%%%%%%%%%%%%%%%%%%%%%%%
%%%%    Weight Function and Casimir Energy Evaluation   %%%%%%
%%%%%%%%%%%%%%%%%%%%%%%%%%%%%%%%%%%%%%%%%%%%%%%%%%%%%%%%%%%%%%%%%%%%%%
\section{
Weight Function and Casimir Energy Evaluation\label{uncert}
}
%***label***{uncert}
We introduce, instead of restricting the integral region, 
a {\it weight function} $W(\ptil,z)$ in the ($\ptil,z$)-space  
for the purpose of suppressing UV and IR divergences of the Casimir Energy. 
%*** uncert1 %%%%%%%%%%%%%%%%
\bea
E^{\mp~W}_{Cas}(\om,T)\equiv\intp\int_{1/\om}^{1/T}dz W(\ptil,z)F^\mp (\ptil,z),
F^\mp(\ptil,z)
=\frac{2}{(\om z)^3}\int_\ptil^\infty\ktil G^\mp_k(z,z)d\ktil     ,\nn
\mbox{Examples}
\left\{
\begin{array}{cc}
(N_1)^{-1}\e^{-(1/2) \ptil^2/\om^2-(1/2) z^2 T^2}\equiv W_1(\ptil,z),\ N_1=1.711/8\pi^2 & \mbox{elliptic}\\
(N_{2})^{-1}\e^{-\ptil zT/\om}\equiv W_2(\ptil,z),\ N_2=2\frac{\om^3}{T^3}/8\pi^2       & \mbox{hyperbolic1}\\
(N_{8})^{-1}\e^{-1/2 (\ptil^2/\om^2+1/z^2T^2)}\equiv W_8(\ptil,z),\ N_8=0.4177/8\pi^2 & \mbox{reciprocal1}\\
\end{array}
           \right.
\label{uncert1}
\eea
%%%%%%%%%%%%%%%%%%%%%%%%%%%%%
where 
$G_k^\mp(z,z) $ are defined in (\ref{HKA12}). 
%$\al, \be, \ga, \del, \ep, \tau$ and $\om$ are some appropriate constants. 
$N_i$'s are the normalization constants. 
We show the shape of the energy integrand $(-1/2)\ptil^3W_1(\ptil,z)F^-(\ptil,z)$ in 
Fig.\ref{PlanckDistBW1L2mank5senT1}b. 
                            %%%      Fig.10 >>  ***    %%%

We can check the divergence (scaling) behavior of $E^{\mp~W}_{Cas}$ by 
{\it numerically} evaluating the $(\ptil,z)$-integral (\ref{uncert1}) for 
the rectangle region of Fig.\ref{zpINTregionWzpINTregionW2}a. \q $E^W_{Cas}=$
%*** uncert1bX %%%%%%%%%%%%%%%%
\bea
\left\{
\begin{array}{cc}
\frac{\om^4}{T}\La\left\{  0.42+0.01~\ln\frac{\La}{\om}+0\times \ln\frac{\La}{T}  \right\} & \mbox{for}\q W_1 \\
\frac{T^2}{\om^2}\La^4 (6.45+0.234 \ln~\frac{\La}{\om}-0.479\ln~\frac{\La}{T})\times 10^{-2}  &\mbox{for}\q W_2\\
\frac{\om^4}{T}\La\left\{  0.42+0.02~\ln\frac{\La}{\om}+0\times \ln\frac{\La}{T}  \right\}
   & \mbox{for}\q W_8
\end{array}
           \right.
\label{uncert1bX}
\eea
%%%%%%%%%%%%%%%%%%%%%%%%%%%%%
%**(MathematicaWork060719/IntegralCalculation0704/CasEneFlatWeighted/)**
%
%The round-bracketed triplet data, corresponding to 3 values of $l$, are not stable. 
%In particular the $\La^n\ln\La$-coefficients decreases as $1/l$ (except $W_{88}$). 
%we take $\al=\be=\ga=\del=\ep=\tau=\om=1$ in the evaluation.
They give, after normalizing the factor $\La/T$, {\it only} the {\it log-divergence}. 
%*** uncert1c %%%%%%%%%%%%%%%%
\bea
E^W_{Cas}/\La T^{-1} =\al \om^4\left( 1-4c\ln (\La/\om) \right) 
\com
\label{uncert1c}
\eea
%%%%%%%%%%%%%%%%%%%%%%%%%%%%%
where $\al$ and $c$ can be read from (\ref{uncert1bX}) depending on the choice of $W$. 
We have confirmed the above result for other many examples\cite{SI0812}. 
This means the 5D Casimir energy is {\it finitely} obtained by the ordinary 
renormalization of the warp factor $\om$. (See the final section.)
%*****As for $c$, we expect it reaches a fixed value as $l$ increases furthermore.***
In the above result of the warped case, the IR parameter $l$ in the flat result
(\ $
E^{W,flat}_{Cas}/\La l =-\frac{\al}{l^4}\left( 1-4c\ln (l\La) \right) =-\frac{\al}{{l'}^4}
$\ )
is replaced by the inverse of the warp factor $\om$.

%%%%%%%%%%%%%%%%%%%%%%%%%%%%  Sec.7  %%%%%%%%%%%%%%%%%%%%%%%%%%%%%%%%%   
%%%%  Definition of Weight Function and Dominant Path     %%%%%%
%%%%%%%%%%%%%%%%%%%%%%%%%%%%%%%%%%%%%%%%%%%%%%%%%%%%%%%%%%%%%%%%%%%%%%
\section{
Meaning of Weight Function and Quantum Fluctuation of Coordinates and Momenta\label{weight}
}
%***label***{weight}
In order to most naturally accomplish 
the above procedure, we can go to a %drastically 
{\it new step}. Namely, 
we {\it propose} to {\it replace} the 5D space integral with the weight $W$, (\ref{uncert1}), 
by the following {\it path-integral}\cite{Fey72}. We 
{\it newly define} the Casimir energy in the higher-dimensional theory as follows.\q  $\Ecal_{Cas}(\om,T,\La)\equiv $
%*** weight5 %%%%%%%%%%%%%%%%
\bea
\int_{\frac{1}{\La}}^{\frac{1}{\m}}d\rho
\int_{
\begin{array}{c}
r(1/\om)=\rho\\
r(1/T)=\rho
\end{array}
      }
\prod_{a,z}\Dcal x^a(z)F(\frac{1}{r},z)
\exp\left[ 
-\frac{1}{2\al'}\int_{1/\om}^{1/T}\frac{1}{\om^4z^4}\sqrt{{r'}^2+1}~r^3 dz
    \right],
\label{weight5}
\eea 
%%%%%%%%%%%%%%%%%%%%%%%%%%%
where $\m=\La T/\om$ and the limit $\La T^{-1}\ra \infty$ is taken. 
The string (surface) tension parameter $1/2\al'$ is introduced. (Note: Dimension of $\al'$ is [Length]$^4$. ) 
The square-bracket ($[ \cdots ]$)-parts of (\ref{weight5}) are \ 
$-\frac{1}{2\al'}$Area = $-\frac{1}{2\al'}\int\sqrt{\mbox{det}g_{ab}}d^4x$ 
 where $g_{ab}$ is the induced metric on the 4D surface. 
$F(\ptil,z)$ is defined in (\ref{HKA13}) and shows 
the {\it field-quantization} of the bulk scalar (EM) fields. 

The proposed definition, (\ref{weight5}), clearly shows the 4D space-coordinates $x^a$ 
or the 4D momentum-coordinates $p^a$ are {\it quantized}
 (quantum-statistically, not field-theoretically) with 
the Euclidean time $z$ and the "{\it area} Hamiltonian" 
$A=\int\sqrt{\det g_{ab}}~d^4x$. 
%\footnote
\ [\small{
We recall the similar situation occurs in the standard string approach. 
The space-time coordinates obey some uncertainty principle\cite{Yoneya87}.  
}]\ 
Note that $F(\ptil,z)$ or $F(1/r,z)$ 
%are difined in (\ref{uncert1}) and 
appears, in (\ref{weight5}), as the {\it energy density operator} in the quantum statistical system of
$\{ p^a(z)\}$ or $\{ x^a(z)\}$. 

In the view of the previous paragraph, the treatment of Sec.\ref{uncert} is an {\it effective} action 
approach using the weight function $W(\ptil,z)$. 
Note that the integral over $(p^\m,z)$-space, appearing in (\ref{HKA13}), 
is the summation over all degrees of freedom of the 5D space(-time) points using the "naive" measure 
$d^4pdz$. 
An important point is that we have the possibility to take another  
measure for the summation. 
We have adopted, in Sec.\ref{uncert}, the new measure $W(p^\m,z)d^4pdz$ in such a way that the Casimir energy 
{\it does not show physical divergences}. 
We expect the direct evaluation of (\ref{weight5}), numerically 
or analytically, leads to the similar result. 
%%%%%%%%%%%%%%%%%%%%%%%%%%%%  Sec.8  %%%%%%%%%%%%%%%%%%%%%%%%%%%%%%%%%   
%%%%  Discussion and Conclusion     %%%%%%
%%%%%%%%%%%%%%%%%%%%%%%%%%%%%%%%%%%%%%%%%%%%%%%%%%%%%%%%%%%%%%%%%%%%%%
\section{
Discussion and Conclusion\label{conc}
}
%***label***{conc}
The log-divergence, (\ref{uncert1c}), is renormalized away by
%*** conc1 %%%%%%%%%%%%%%%%
\bea
E^W_{Cas}/\La T^{-1} 
%=-\al \om^4\left( 1-4c\ln (\La/\om) \right) 
=\al \om'^4\com\q 
\om'=\om\sqrt[4]{1-4c\ln (\La/\om) }\pr
\label{conc1}
\eea
%%%%%%%%%%%%%%%%%%%%%%%%%%%%%
Taking into account the fact $|c|\ll 1$, 
we find the renormalization group function for the warp factor $\om$ as  
%*** conc2 %%%%%%%%%%%%%%%%
\bea
|c|\ll 1\com\q
\om'=\om (1-c\ln (\La/\om) )\com\q
\be (\mbox{$\be$-function})\equiv \frac{\pl}{\pl(\ln \La)}\ln \frac{\om'}{\om}=-c
\pr
\label{conc2}
\eea
%%%%%%%%%%%%%%%%%%%%%%%%%%%%%
We should notice that, in the flat geometry case, the IR parameter (extra-space size) $l$ 
is renormalized . In the present warped case, however, the corresponding parameter $T$ 
is {\it not renormalized}, but the {\it warp parameter} $\om$ {\it is renormalized}. 
Depending on the sign of $c$, the 5D bulk curvature $\om$ 
{\it flows} as follows. When $c>0$, the bulk curvature $\om$ decreases (increases) as the 
the measurement energy scale $\La$ increases (decreases). 
When $c<0$, the flow goes in the opposite way. 

Recently the dark energy ( as well as the dark matter ) in the universe is a hot subject. 
It is well-known that the dominant candidate is the cosmological term. 
We also know the proto-type higher-dimensional theory, that is, the 5D KK theory, 
has predicted so far the {\it divergent} cosmological constant\cite{AC83}. 
This unpleasant situation has been annoying us for a long time. 
If we apply the present result, the situation drastically improve. The cosmological
constant $\la$ appears as
%*** conc3%%%%%%%%%%%%%%%%
\bea
S=\int d^4x \sqrt{-g}\{\frac{1}{G_N}(R+\la) \}
+\int d^4x \sqrt{-g}\{\Lcal_{mat}\},
R_\mn-\half g_\mn R-\la g_\mn =T_\mn^{mat},
\label{conc3}
\eea
%%%%%%%%%%%%%%%%%%%%%%%%%%%%%
where $G_N$ is the Newton's gravitational constant, $R$ is the Riemann scalar 
curvature. The constant $\la$ observationally takes the value:\ 
%*** conc4%%%%%%%%%%%%%%%%
%\bea
$
\frac{1}{G_N}\la_{obs}\sim \frac{1}{G_N{R_{cos}}^2}\sim m_\n^4\sim (10^{-3} eV)^4
,
$
%\label{conc4}
%\eea
%%%%%%%%%%%%%%%%%%%%%%%%%%%%%
\ where $R_{cos}$ is the cosmological size (Hubble length), $m_\n$ is the neutrino mass.
%\footnote
\ [\small{ 
The relation $m_\n\sim \sqrt{M_{pl}/R_{cos}}=\sqrt{1/R_{cos}\sqrt{G_N}}$ appears  
in some extra dimension model\cite{SI0012,SI01Tohwa}. The neutrino mass is located 
at the geometrical average of two extreme ends of the mass scales in the universe.  
}]\  
On the other hand, we theoretically so far have the value:\ 
%*** conc5%%%%%%%%%%%%%%%%
%\bea
$
\frac{1}{G_N}\la_{th}\sim \frac{1}{{G_N}^2}={M_{pl}}^4\sim (10^{28} eV)^4
\ .
$
%\label{conc5}
%\eea
%%%%%%%%%%%%%%%%%%%%%%%%%%%%%
\ This is because the mass scale usually comes from the quantum 
gravity. (See Ref.~\refcite{SI83NPB} for the derivation using the  
Coleman-Weinberg mechanism.) 
We have the famous huge discrepancy factor 
$\la_{th}/\la_{obs}\sim 10^{124}$ which appears in the Dirac's large 
number theory\cite{PD78}. 
If we apply the present approach, 
we have the warp factor $\om$, and the result (\ref{conc1}) strongly 
suggests the following choice:
%*** conc6%%%%%%%%%%%%%%%%
\bea
\frac{1}{G_N}{\tilde \la}_{th}=\al {\om'}^4\ ,\ 
\om'\sim
 \frac{1}{\sqrt[4]{G_N{R_{cos}}^2}}=\sqrt{\frac{M_{pl}}{R_{cos}}}\sim m_\n\sim 10^{-3}\mbox{eV}\ ,\ 
{\tilde \la}_{th}\sim \frac{1}{{R_{cos}}^2}
.
\label{conc6}
\eea
%%%%%%%%%%%%%%%%%%%%%%%%%%%%% 
Note that this choice $\om'$ is within the allowed region obtained from 
the Newton force test (see Ref.~\refcite{SI01CQG}). 
We succeed in obtaining  the {\it finite} cosmological constant and 
its gross absolute value consistent with the observed one.  
Now we can understand that the {\it smallness of the cosmological constant comes from 
the renormalization flow} for the non asymptotic-free case ($c<0$ in (\ref{conc2})). 
%\footnote
\ [\small{
In the 2 dim R$^2$-gravity, the same thing occurs\cite{SI95NPB}. 
}]\ 
In this case the choice of the regularization parameters are
%*** conc7%%%%%%%%%%%%%%%%
\bea
T\sim 10^{-33}\ \mbox{eV}((\mbox{Cosm. Size})^{-1}, 1/R_{cos})\com\q
\La\sim 10^{28}\ \mbox{eV}(\mbox{Planck mass}, M_{pl})
\com
\label{conc7}
\eea
%%%%%%%%%%%%%%%%%%%%%%%%%%%%%
and the relations, $\om\sim \sqrt{T\La}$ and $\mu=\La T/\om \sim \sqrt{T\La}$ are valid. 
The normalization factor in (\ref{uncert1c}) is given by 
%*** conc8%%%%%%%%%%%%%%%%
%\bea
$
\frac{\La}{T}\sim 10^{61}
.
$
%\label{conc8}
%\eea
%%%%%%%%%%%%%%%%%%%%%%%%%%%%%
\ The total number of the unit spheres within the large sphere is given by 
%*** conc9%%%%%%%%%%%%%%%%
%\bea
$
\frac{\La^4}{\m^4}=\frac{\om^4}{T^4}\sim 10^{120}
,
$
%\label{conc9}
%\eea
%%%%%%%%%%%%%%%%%%%%%%%%%%%%%
which is expected to show 
the degree of freedom of the universe (4D space-time).

\end{document}